\documentclass[aps,prl,article]{revtex4}

\newcommand{\Sp}{{\rm Sp}\hspace{0.1em}}

\begin{document}

\title{\bf Quantum Monte Carlo determinantal algorithm
without Hubbard-Stratonovich transformation: a general
consideration}
\author{A.N. Rubtsov}
\email{alex@shg.ru}
\affiliation{Physics Department, Moscow State
University, 119992 Moscow, Russia}

\begin{abstract}
    Continuous-time determinantal algorithm is proposed for the quantum Monte Carlo
    simulation of the interacting fermions. The scheme does not
    invoke Hubbard-Stratonovich transformation. The fermionic
    action is divided into two parts. One of them contains the interaction
    and certain additional terms; another one is purely Gaussian.
    The first part is considered as a perturbation.
    Terms of the series expansion for the partition function are
    generated in a random walk process. The sign problem and
    the complexity of the algorithm are analyzed. We argue that the
    scheme should be useful particularly for the systems with non-local interaction.
\end{abstract}

\maketitle

    The family of quantum Monte Carlo (QMC) methods is the most
universal tool for the numerical study of quantum many-body
systems with strong correlations. So-called determinental methods
\cite{Scalapino, Hirsh} are commonly used For the interacting
fermions, because other known QMC schemes (like Stochastic Series
Expansion \cite{SSE} or worm algorithms \cite{Worm}) suffer an
unacceptably bad sign problem for this case. The partition
function of the system is presented as a sum over an ensemble of
the systems of uncoupled fermions. This idea was proposed by
Scalapino and Sugar \cite{Scalapino}. Their algorithm was later
significantly improved by Hirsh \cite{Hirsh}. The scheme by Hirsh
is now almost standard for QMC simulations. Two points are
essential for the method: first, the imaginary time is
artificially discretized, and then a discrete Hubbard-Stratonovich
transformation is performed at each time slice to decouple the
fermionic degrees of freedom. After this decoupling, the fermionic
degrees of freedom can be integrated out, and Monte Carlo sampling
should be performed in the space of auxiliary Hubbard-Stratonovich
spins. For a system of $N$ atoms the number of spins scales as
$\beta N /\tau_0$ for the local (short-range) interaction and as
$\beta N^2/\tau_0$ for the long-range one ($\beta$ is the inverse
temperature, and $\tau_0$ is a time slice).

One can point out the following drawbacks of the Hirsh algorithm.
First, for large systems and small temperature, the sign problem
results in the exponential growth of computational time. This is a
principal problem of QMC, and there is no solution on the horizon
now. However, for relatively small clusters, and in particular for
dynamical mean-field calculations \cite{DMFT}, the sign problem is
not of crucial importance. Further, the non-locality of
interaction hampers the calculation,  because it is hard to
simulate systems with a large number of auxiliary spins. It is
nearly impossible to simulate the system with interactions that
are non-local in time, when the number of spins is proportional to
$(\beta/\tau_0)^2$. And finally, the time discretization leads in
a systematic error of the result.

In 1999th a continuous-time modification of the algorithm was
proposed \cite{CT}. It was based on a series expansion for the
partition function in powers of interaction. The scheme is free of
systematic errors. The Hubbard-Stratonovich transformation is
still invoked. The number of auxiliary spins scales similarly to
the discrete scheme.

In this communication we present a continuous-time algorithm which
is based on the series expansion as well but {\it does not} deal
with any auxiliary-field variables. For the case of Hubbard model
the sign problem and the computational time are found to be
similar to what occurs for the Hirsh scheme. The principal
advantages of the present algorithm are related to the different
scaling of the computational time for non-local interactions. The
paper is aimed at a general description of the algorithm and the
estimation of the computation time.  Our pilot numerical results
are not presented  here because the systematic study is not yet
done.

Consider a system of fermions with pair interaction. For the most
general situation, the partition function can be presented as
follows:
\begin{eqnarray}\label{Z0}
    &Z={\Sp} T e^{-S}\\  \nonumber
    &S=\int \int   t_r^{r'}  c^\dag_{r'} c^{r} dr  dr'+ \int \int \int \int
    w_{r_1 r_2} ^{r_1' r_2'} c^\dag_{r_1'} c^{r_1} c^\dag_{r_2'} c^{r_2} dr_1 dr_1' dr_2 dr_2'.
\end{eqnarray}
Here $T$ is a time-ordering operator, $r=\{\tau,s,i\}$ is a
combination of the continuous imaginary-time variable $\tau$, spin
orientation $s$ and the discrete index $i$ numbering the
single-particle states in a lattice. For the single-band models
$i$ is an atom number. For electrons $s=\uparrow$ or $\downarrow$.
Integration over $d r$ implies the integral over $\tau$, and the
sum over all lattice states and spin projections: $\int dr \equiv
\sum_i \sum_s \int_0^\beta d\tau$.

We borrow the linear-algebra style for sub- and superscripts to
make the notation clearer. The creation and annihilation operators
are labelled as covariant and contravariant vectors, respectively.
The labelling for coefficients $t, w$ is chosen to present all
integrands like scalar products of tensors. In principle, we could
declare the summation over the repeating indices and would not
explicitly write integrals. However, we keep them for convenience.

Let us introduce an additional quantity $\alpha_{r'}^{r}$, and
split $S$ into two parts. Up to an additive constant
\begin{eqnarray}\label{Z1}
    &S=S_0+W,\\  \nonumber
    &S_0=\int \int   \left( t_r^{r'}+\int \int
\alpha^{r_2}_{r_2'} (w_{r r_2} ^{r' r_2'} + w_{r_2 r} ^{r_2' r'})
dr_2 dr_2' \right) ~ c^\dag_{r'} c^{r} ~ dr  dr',\\  \nonumber
    &W=\int \int \int \int
    w_{r_1 r_2} ^{r_1' r_2'} (c^\dag_{r_1'} c^{r_1}-\alpha^{r_1}_{r_1'})
    (c^\dag_{r_2'} c^{r_2}-\alpha^{r_2}_{r_2'}) dr_1 dr_1' dr_2 dr_2'.
\end{eqnarray}
The freedom to choice $\alpha_{r'}^{r}$ is to be used later to
optimize the algorithm, particularly to minimize the sign problem.

We consider $S_0$ as an unperturbed action and switch to the
interaction representation. The perturbation-series expansion for
$Z$ is as follows:
\begin{eqnarray}\label{ser}
    &Z=\sum_{k=0}^\infty \int dr_1 \int d r_1' ... \int dr_{2k} \int
    dr_{2k}' \Omega_k (r_1, r_1', ..., r_{2k}, r_{2k}')\\
     \nonumber
     &\Omega_k=Z_0 \frac{(-1)^k}{k!}w_{r_1 r_2}^{r_1' r_2'} \cdot ... \cdot w_{r_{2k-1} r_{2k}}^{r_{2k-1}' r_{2k}'}
     D^{r_1 r_2 ... r_{2k}}_{r_1' r_2' ... r_{2k}'}
\end{eqnarray}
where $Z_0$ is a partition function for the unperturbed system and
\begin{equation}\label{D}
    D^{r_1 ... r_{2k}}_{r_1' ... r_{2k}'}
    = <T (c^\dag_{r_1'} c^{r_1}-\alpha^{r_1}_{r_1'}) \cdot ... \cdot (c^\dag_{r_{2k}'} c^{r_{2k}}-\alpha^{r_{2k}}_{r_{2k}'})>.
\end{equation}
Hereafter the triangle brackets denote the average over the
unperturbed system, $<A>=Z_0^{-1} \Sp (T A e^{-S_0})$.

To obtain an expression for $D$, one can start from the case of
$\alpha=0$. Since $S_0$ is Gaussian, Wick theorem applies.
Therefore D is a determinant $2k \times 2k$ matrix. Two-point
Green functions
\begin{equation}
    g_{r'}^r=<T c^\dag_{r'} c^{r}>
\end{equation}
form this matrix for $\alpha=0$. Now, the non-zero $\alpha$ can be
taken into account. It is easy to prove that
\begin{equation}\label{det}
    D^{r_1 r_2 ... r_{2k}}_{r_1' r_2' ... r_{2k}'}=
    \det|| g^{r_i}_{r_j'}-\delta(i-j) \alpha^{r_i}_{r_j'} ||
\end{equation}
where $\delta(i-j)$ is a delta-symbol.

Like any other other QMC scheme, the proposed one is based on a
Markov process. Points in the configuration space are determined
 by the number $k$ and the set $r_1, r_1', ..., r_{2k}, r_{2k}'$. Suppose for a
moment that $\Omega$ is always positive, and consider a random
walk with a probability of $Z^{-1} \Omega_k(r_1, r_1', ...,
r_{2k}, r_{2k}')$ to visit each point.
Then for example the Green
function can be calculated from
\begin{eqnarray}\label{G}
    &Z^{-1} <T c^\dag_{r'} c^{r} e^{-W}>=\overline{g_{r'}^r (r_1, r_1', ..., r_{2k}')}
    \equiv\\  \nonumber
    &\equiv\sum_k \int dr_1 \int d r_1' ... \int dr_{2k}
    g_{r'}^r (r_1, r_1', ..., r_{2k}') \Omega_k(r_1, r_1', ...,
    r_{2k}'),
\end{eqnarray}
where $g_{r'}^r(r_1, r_1', ..., r_{2k}')$ determines the Green
function for a current realization
\begin{equation}
    g_{r'}^r(r_1, r_1', ..., r_{2k}')=(D^{r_1 ... r_{2k}}_{r_1' ... r_{2k}'})^{-1} <T c^\dag_{r'} c^{r}
    (c^\dag_{r_1'} c^{r_1}-\alpha_{r'_1}^{r_1})\cdot...\cdot(c^\dag_{r_1'}
    c^{r_1}-\alpha_{r'_1}^{r_1})>.
\end{equation}
The overline stands for the averaging over the above-mentioned
random walk. The important notice is that the series expansion for
an exponent {\it always} converges, therefore the discussed
averaging is always well-defined in a mathematical sense.

The sign problem should be discussed first. If $\Omega$ is not
always positive, one should work with a probability $|\Omega|$ and
calculate $\overline{g \rm{sign} (\Omega)}/\overline{ \rm{sign}
(\Omega)} $ instead of $\overline{g}$. For a practical calculation
it is desirable to maximize the average sign, because otherwise a
computational error bar is unacceptable.

A proper choice of $\alpha$ can suppress the sign problem in
certain cases. To be concrete, let us consider a Hubbard model. In
this model the interaction is local in time and space, and only
electrons with opposite spins interact. Therefore it is reasonable
to take $\alpha_{t' i' \uparrow}^{t i \uparrow}
=\delta(\tau-\tau') \delta(i-i') \alpha_{\uparrow}$, the similar
for $\alpha_\downarrow$, and
$\alpha^\downarrow_\uparrow=\alpha_\downarrow^\uparrow=0$. The
perturbation $W$ becomes
\begin{equation}\label{Hubbard}
W_{Hubbard}=U \int
(n_\uparrow(\tau)-\alpha_\uparrow)(n_\downarrow(\tau)-\alpha_\downarrow)
dt
\end{equation}
Here Hubbard $U$ and the occupation number operator $n=c^\dag c$
are introduced. The Gaussian part of Hubbard action is
spin-independent and does not rotate spins. This means that only
$g_\downarrow^\downarrow, g^\uparrow_\uparrow$ do not vanish, and
the determinant in (\ref{det}) is factorized
\begin{equation}\label{updown}
D^{r_1 r_2 ... r_{2k}}_{r_1' r_2' ... r_{2k}'}=D^{r_1 r_3 ...
r_{2k-1}}_{r_1' r_3' ... r_{2k-1}'} D^{r_2 r_4 ... r_{2k}}_{r_2'
r_4' ... r_{2k}'} \equiv D_\uparrow D_\downarrow
\end{equation}
For the case of attraction $U<0$ one should choose
\begin{equation}\label{attract}
    \alpha_\uparrow=\alpha_\downarrow=\alpha,
\end{equation}
where $\alpha$ is a real number. For this choice
$g_\downarrow^\downarrow=g^\uparrow_\uparrow$, and therefore
$D_\uparrow=D_\downarrow$. $\Omega$ is always positive in this
case, as it follows from formula (\ref{ser}).

This choice of $\alpha$ is useless for a system with repulsion,
however. Compared to the case of attraction, another sign of $w$
at $\alpha_\uparrow=\alpha_\downarrow$ results in the alternating
signs of $\Omega_k$ with odd and even $k$ \cite{Alt}. Another
condition for $\alpha$ is required. The particle-hole symmetry can
be exploited for the Hubbard model at half-filling. In this case,
the transformation $c^\dag_\downarrow \to \tilde{c}_\downarrow$
converses the Hamiltonian with repulsion to the same but with
attraction. Therefore the series (\ref{ser}) in powers of $W=U
\int (n_\uparrow(\tau)-\alpha)(\tilde{n}_\downarrow(\tau)-\alpha)
dt$ does not contain negative numbers, in accordance to the
previous paragraph. The value of trace in (\ref{ser}) is
independent of a particular representation. In the original
(untransformed) basis the above written $W$ reads as $U \int
(n_\uparrow(\tau)-\alpha) (n_\uparrow(\tau)-1+\alpha) dt$. We
conclude that
\begin{equation}\label{repuls}
\alpha_\uparrow=1-\alpha_\downarrow=\alpha
\end{equation}
eliminates the sign problem for repulsive systems with a
particle-hole symmetry.

Of course, the average sign for a system with repulsion is not
equal to unity in a general case. As usual the exponential
fall-off occurs for the large systems or small temperature.

Now we discuss how to organize a random walk in practice. We need
to perform a random walk in the space of $k; r_1, r_1', ...,
r_{2k}, r_{2k}'$. Two kinds of trial steps are necessary: one
should try either to increase or to decrease $k$ by 1, and,
respectively, to add or to remove the four corresponding
"coordinates". A proposition for $r_{2k+1}, r'_{2k+1}, r_{2k+2},
r'_{2k+2}$ should be generated for the "incremental" step. The
normalized modulus
\begin{eqnarray}
&||w||^{-1} |w_{r_{2k+1} r_{2k+2}}^{r_{2k+1}' r_{2k+2}'}| \\
\nonumber &||w||={\int \int \int \int  |w_{r R}^{r' R'}|} dr dR
dr' dR'
\end{eqnarray}
can be used as a probability density for this proposition. Then
the standard Metropolis acceptance criterion can be constructed
using the ratio
\begin{equation}\label{prob}
    \frac{||w||}{k+1} \cdot
    \left| \frac{D^{r_1  ... r_{2k+2}}_{r_1' ... r_{2k+2}'} }
    {D^{r_1  ... r_{2k}}_{r_1' ... r_{2k}'} } \right|.
\end{equation}

For the "decremental" step one should take a random integer $j$
between 0 and $k-1$ and consider a probability of the removing
$r_{2j+1}, r'_{2j+1}, r_{2j+2}, r'_{2j+2}$. The detailed balance
condition should be satisfied, therefore an inverse of
(\ref{prob}) should be used in the acceptance criterion.

It is possible to introduce more complicated trial steps in a
similar way. It can be an increment or decrement of $k$ by 2,
shifts of $r$ {\it etc}. Normally, the basic steps discussed above
should be sufficient. We shall demonstrate however that for
certain cases the increments (decrements) by 2 are required.

The most time consuming operation of the algorithm is a
calculation of the ratio of determinants. This task is relatively
easy if Gaussian part of action is described by a Hamiltonian
$S_0=\int d\tau H_0(\tau)$ and the interaction is also a
combination of the local in time parts:
\begin{equation}
    w_{r_1 r_2} ^{r_1' r_2'} \propto \delta(\tau_1-\tau_1')
    \delta(\tau_2-\tau_2').
\end{equation}
One can call this system "almost Hamiltonian". Purely Hamiltonian
systems and particularly Hubbard model also belong to this class,
of course. For this case the expression (\ref{D}) for $D$  can be
written explicitly in the Schrodinger representation:
\begin{equation}\label{Schred}
    D^{r_1 ... r_{2k}}_{r_1' ... r_{2k}'}
    = Z_0^{-1}\Sp \left(e^{-\tau_1 H_0} (c^\dag_{i_1'} c^{i_1}-\alpha^{i_1}_{i_1'})
    e^{(\tau_1-\tau_2) H_0}
    \cdot ... \cdot (c^\dag_{i_{2 k'}} c^{i_{2 k}}-\alpha^{i_{2 k}}_{i_{2 k}'})
    e^{ \tau_{2 k} H_0}\right),
\end{equation}
where $\tau_1...\tau_{2k}$ are pre-ordered in time. Further, an
identity $c^\dag_{i'} c^{i}-\alpha=-\alpha \exp(\xi c^\dag_{i'}
c^{i})$ holds with $\xi=-\alpha^{-1}$ for $i \neq i'$ and $\xi=\ln
(1-\alpha^{-1})$ for $i=i'$. Trace here can be calculated in a
Scalapino-Sugar manner \cite{Scalapino}, that requires the
calculation of $\propto N \times N$ matrix determinant for a
system of $N$ atoms. Although the calculation of a determinant
from scratch takes $\propto N^3$ operations, the fast-update
technique can be used here, resulting in a $\propto N^2$
operation-count.

For the general situation, formula (\ref{det}) is to be used to
estimate $D$. Here we present an estimation for the typical value
of $k$. An average value of (\ref{prob}) determines an acceptance
rate for QMC sampling. It is reasonable to expect that by the
order of magnitude this rate is not much less then unity, and
consequently
\begin{equation}
    k \approx ||w|| \overline{\left| \frac{D^{r_1  ... r_{2k+2}}_{r_1' ... r_{2k+2}'} }
    {D^{r_1  ... r_{2k}}_{r_1' ... r_{2k}'} } \right|}
\end{equation}
If the sign problem is absent and the sign of $w_{r_{1}
r_{2}}^{r_{1}' r_{2}'}$ is the same for all arguments, the
right-hand side can be interpreted as an expectation value of
$|W|$ (compare with formula (\ref{G})), {\it i.e.}
\begin{equation}
    k \approx \overline {|W|}.
\end{equation}
For the Hubbard lattice of $N$ atoms, for instance, $|W| \propto
\beta |U| N$. Using fast updates one can make a step in $\propto
k^2$ operations.

To obtain the fastest procedure one needs to minimize $k$ with
respect to $\alpha$'s. It should decrease not only the
computational time for a single random-move, but the
autocorrelation length of the random walk as well. Therefore the
minimization of $k$ is desirable for the almost-Hamiltonian
systems also.

For example, once the the condition (\ref{repuls}) for the Hubbard
model with attraction is fulfilled, the value of $\alpha$ still
can be adjusted. We obtain
\begin{equation}
    \overline {|W|} \propto   \overline{n_\uparrow(\tau)
    n_\downarrow(\tau)}-  2 \overline{n} \alpha + \alpha^2 ,
\end{equation}
where $\overline{n}$ is the average filling factor. Minimization
gives a physically reasonable suggestion $\alpha=\overline{n}$.

It is useful to analyze a toy single-atom Hubbard model to get a
filling of the behavior of series (\ref{ser}). The two parts of
action are
\begin{eqnarray}
&S_0=\int  (-\mu+U \alpha_\downarrow)
n_\uparrow(\tau)+(-\mu+U \alpha_\uparrow) n_\downarrow(\tau)) d\tau;\\
\nonumber &W=U \int (n_\downarrow(\tau)-\alpha_\downarrow)
(n_\uparrow(\tau)-\alpha_\uparrow) d\tau.
\end{eqnarray}
Here $\mu$ is a chemical potential. For a half-filled system
$\mu=U/2$. For this model, it is easy to calculate traces in
(\ref{Schred}), and (\ref{ser}) becomes
\begin{equation}
\Omega_k= \frac{(-U \alpha_\uparrow
\alpha_\downarrow)^k}{k!}\left(1+e^{\beta (\mu-U
\alpha_\downarrow)}(1-\alpha_\uparrow^{-1})^k\right)
        \left(1+e^{\beta (\mu-U \alpha_\uparrow)}(1-\alpha_\downarrow^{-1})^k\right)
\end{equation}
Consider a case of repulsion ($U>0$). Let us use a condition
(\ref{repuls}) for an arbitrary filling factor. The later
expression can be presented in the form
\begin{equation}\label{imp}
\Omega_k=e^{\beta (\mu-U \alpha)} \frac{(U \alpha^2)^k}{k!}
\left(1+e^{\beta (\mu-U+U
\alpha)}(1-\alpha^{-1})^k\right)
        \left(1+e^{\beta (-\mu+U \alpha)}(1-\alpha^{-1})^k\right)
\end{equation}
For $\mu=U/2$ the value of $\Omega_k$ is positive for any
$\alpha$. For a general filling factor, the situation depends on
the value of $\alpha$. For $0<\alpha<1$ negative numbers can occur
at certain $k$. Outside this interval all terms are positive, and
there is no sign problem for a single-atom system under
consideration.

 Since  the sign problem persists already
for the impurity problem for $0<\alpha<1$, such a choice is also
not suitable  for the $N$-atom repulsive Hubbard system. On the
other hand, minimization of $\bar{W}$ requires $\alpha$ to be as
close to this interval as possible. Therefore it is reasonable to
take $\alpha=1$ or slightly above. This is the same as zero or a
small negative value, since $\alpha_\uparrow=1-\alpha_\downarrow$.
In this case our pilot calculations and the discrete-time scheme
\cite{Hirsh} demonstrate approximately the same sign problem.

There is also an important note about the system at half-filling.
For this case the optimal choice is
$\alpha_\uparrow=\alpha_\downarrow=1/2$. In this case one can
observe from (\ref{imp}) that $\Omega_k=0$ for any odd $k$. This
is a consequence of a particle-hall symmetry. This is an example
of the situation when the trial increase/decrese of $k$ by 1 does
not help, and it is necessary to consider changes of $k$ by 2.

In a conclusive part, we would like to discuss possible benefits
of the proposed algorithm over another determinantal schemes. We
demonstrated that for a Hubbard-type models the computational time
for a single trial step scales as $(\beta U N)^2$ for the general
case and as  $N^2$ for a Hamiltonian system. This is the same
scaling as for the schemes based on a Stratonovich transformation.
An important difference occurs however for the non-local
interactions. Consider, for example, a system with a large Hubbard
$U$ and much smaller but still important Coulomb interatomic
interaction. One needs to introduce $N^2$ auxiliary fields per
time slice instead of $N$ to take the long-range forces into
account. This slows the calculation dramatically. On the other
hand, the complexity of the present algorithm should remain almost
the same as for the local interactions, because $\overline{|W|}$
does not change much. It should be also possible to study the
interactions retarded in time, particularly the effects related to
dissipation.

The proposed algorithm operates in a continuous time and does not
invoke Statonovich transformation. A methodologist can ask if
there is a similar scheme for a discrete time. The answer is yes,
one should only replace integrals with sums in all formulae. The
time-scaling should remain the same as for the continuous time.
The discrete-time formulation might be handier for the practical
realization.

I am grateful to A.I. Lichtenstein and F. Assaad for their very
valuable comments. This research was supported in part by the
National Science Foundation under Grant No. PHY99-07949.

\end{document}